\documentclass[paper]{JHEP}
\usepackage{epsfig}
\def\combi#1#2{\left(\begin{array}{c}#1\\#2\end{array}\right)}
\def\as{\alpha_{\mbox{\tiny S}}}
\def\alb{\bar\as}
\def\cG{{\cal G}}
\def\gE{\gamma_{\mbox{\tiny E}}}
\def\DL{{\mbox{\scriptsize DL}}}
\def\SL{{\mbox{\scriptsize SL}}}
\def\mR{\mu_{\mbox{\tiny R}}}
\def\res{\,{\mbox{\scriptsize jet}}}
\def\om{\omega}
\def\alom{\frac{\alb}{\om}}
\def\glip{\gamma_{\mbox{\tiny L}}(\alb/\om)}
\title{\boldmath Jet Rates at Small $x$ to Single-Logarithmic Accuracy
\footnote{Work supported in part by the UK Particle Physics and
Astronomy Research Council and by the EU Fourth Framework Programme
`Training and Mobility of Researchers', Network `Quantum Chromodynamics
and the Deep Structure of Elementary Particles',
contract FMRX-CT98-0194 (DG 12 - MIHT).}}
\author{Carlo Ewerz$^{a,b}$ and Bryan R. Webber$^a$\\
$^a$Cavendish Laboratory, University of Cambridge,\\
Madingley Road, Cambridge CB3 0HE, U.K.\\[1mm]
$^b$DAMTP, University of Cambridge,\\
Silver Street, Cambridge CB3 9EW, U.K.\\[1mm]
E-mail: \email{carlo@hep.phy.cam.ac.uk}, \email{webber@hep.phy.cam.ac.uk}}
\abstract{We present predictions of jet rates in deep inelastic
scattering at small $x$ to leading-logarithmic order in $x$,
including all sub-leading logarithms of $Q^2/\mR^2$ where $\mR$
is the transverse momentum scale at which jets are resolved.
We give explicit results for up to three jets, and a perturbative
expansion for multi-jet rates and jet multiplicities.}
\keywords{Deep Inelastic Scattering, QCD, Jets}
\preprint{Cavendish--HEP--99/02\\DAMTP--1999--44\\hep-ph/9904244}
\begin{document}
\section{Introduction}
The summation of logarithms of $1/x$ in deep inelastic structure functions
at small values of Bjorken $x$ leads to the Balitskii-Fadin-Kuraev-Lipatov
(BFKL) equation \cite{BFKL,FR}, which in the leading approximation sums
terms of order $[\as\ln(1/x)]^n$. Recently the next-to-leading terms
have also been computed \cite{NLOA,NLOB}.

Since the dynamics of the small-$x$ region is supposed to be different
from that at higher $x$ in several respects, it is important to make and
test predictions of a wide range of observables in this region.
In recent papers \cite{FSV,Web98} predictions were presented for
the rates of emission of fixed numbers of `resolved' final-state gluons,
together with any number of unresolvable ones. Here `resolved'
means having a transverse momentum larger than some fixed value $\mR$.
The predictions were valid in the double-logarithmic (DL) approximation,
i.e.\  retaining only terms of the form $[\as\ln(1/x)\ln(Q^2/\mR^2)]^n$.
In this approximation, each resolved gluon
can be equated to a single jet, since to resolve it into more than
one jet would cost extra powers of $\as$ with no corresponding powers
of $\ln(1/x)$.

The present paper extends the work of ref.~\cite{Web98} to include terms
with fewer powers of $\ln(Q^2/\mR^2)$, i.e.\ those of the form
$[\as\ln(1/x)]^n\,[\ln(Q^2/\mR^2)]^m$ where $0<m<n$, which we refer
to as single-logarithmic (SL) corrections. The fact that we still
demand a factor of $\ln(1/x)$ with each power of $\as$ means that
the identification of resolved gluons with jets\footnote{When counting
jets in deep inelastic lepton scattering, we always omit final-state
hadrons that originate from the quark-antiquark pair which couples
the gluon to the virtual photon.} remains valid.

There are two alternative methods for the calculation of final-state
properties at small $x$: the original multi-Regge BFKL method
and the CCFM \cite{C,CFM,M} approach, which takes
account of the coherence of soft gluon emission. It has been shown,
at the DL level in refs.~\cite{FSV,Web98} and now at the SL level
\cite{Sal99}, that the two methods are equivalent for the
observables considered here.  We therefore adopt the BFKL approach,
which is calculationally simpler.

The paper is organized as follows. In sect.~\ref{sec_bfkl}
we recall the BFKL formalism and the predicted behaviour of the
gluon structure function at small $x$. In sect.~\ref{sec_rates} we
first compute the single-jet rate to SL accuracy and show that the only
modification to the DL result comes from the SL corrections
to the anomalous dimension. Next, in subsect.~\ref{sec_2jet},
we calculate the SL corrections
to the two-jet rate, first in the form of a
numerical integral and then as a perturbation series.
In subsect.~\ref{sec_3jet} we apply the same methods to the three-jet
rate. In sect.~\ref{sec_njet} we derive the SL perturbative expansion
of the generating function for the multi-jet rates, and use
this to obtain the corresponding expansions for the
mean jet multiplicity and its dispersion.  Our conclusions
are presented in sect.~\ref{sec_conc}. A useful class of
integrals is evaluated to SL accuracy in the appendix.

\section{BFKL formalism}\label{sec_bfkl}
We start from the unintegrated structure function of a single gluon,
$f(x,k^2,\mu^2)$, which in the exclusive form of the BFKL
approach satisfies the equation
\begin{equation}\label{fxk2}
f(x,k^2,\mu^2) = \delta(1-x)\,\delta^2(k) +
\alb\int_{\mu^2}\frac{dq^2}{q^2}\frac{d\phi}{2\pi}
\frac{dz}{z^2}\Delta(z,k^2,\mu^2)\,f(x/z,|q+k|^2,\mu^2)\;.
\end{equation}
Here $\alb=3\as/\pi$, $k$ is the (2-vector) transverse momentum of the
gluon probed in the deep inelastic scattering, $q$ is that of an emitted
gluon, $\phi$ is the azimuthal angle giving the direction of $q$,
$\mu$ is a collinear cutoff and $\Delta$ is the Regge form factor
\begin{equation}
\Delta(z,k^2,\mu^2) =
\exp\left(-\alb\ln\frac{1}{z}\ln\frac{k^2}{\mu^2}\right)\;.
\end{equation}
To carry out the $z$ integration it is convenient to use a Mellin
representation,
\begin{equation}\label{mellin}
f_\om(k^2,\mu^2) = \int_0^1 dx\,x^\om f(x,k^2,\mu^2)\;,
\end{equation}
with inverse
\begin{equation}\label{melinv}
f(x,k^2,\mu^2) = \frac{1}{2\pi i}\int_C d\om\,x^{-\om-1}
f_\om(k^2,\mu^2)\;,
\end{equation}
where the contour $C$ is parallel to the imaginary axis and to the
right of all singularities of the integrand.  This gives
\begin{equation}\label{bfkl}
f_\om(k^2,\mu^2) = \delta^2(k) + H_\om(k^2,\mu^2)
\int_{\mu^2}\frac{dq^2}{q^2}\frac{d\phi}{2\pi}f_\om(|q+k|^2,\mu^2)
\end{equation}
where
\begin{equation}\label{delom}
H_\om(k^2,\mu^2)= \frac{\alb}{\om+\alb\ln(k^2/\mu^2)}\;.
\end{equation}
To solve eq.~(\ref{bfkl}) one can make use of the relation, derived in
the appendix,
\begin{equation}\label{intf}
\int_{\mu^2}^{Q^2}\frac{dq^2}{q^2}\frac{d\phi}{2\pi}f(|q+k|^2) =
\int_{\mu^2}^{Q^2}\frac{dq^2}{q^2}f(\max\{q^2,k^2\})
+2\sum_{m=1}^\infty \zeta(2m+1)
\left(k^2\frac{\partial}{\partial k^2}\right)^{2m}f(k^2)\;,
\end{equation}
which is valid for $\mu^2<k^2<Q^2$ to logarithmic accuracy,
i.e.\ neglecting terms suppressed by powers of $\mu^2/k^2$ or $k^2/Q^2$.
Then for $k^2>\mu^2$ the solution is of the form
\begin{equation}\label{fomsol}
f_\om(k^2,\mu^2) = \frac{\gamma}{\pi k^2}\left(\frac{k^2}{\mu^2}\right)^\gamma
\end{equation}
where
\begin{equation}
1=H_\om(k^2,\mu^2)\left(\ln\frac{k^2}{\mu^2}-\frac{1}{\gamma-1}
+2\sum_{m=1}^\infty \zeta(2m+1)(\gamma-1)^{2m}\right)
\end{equation}
and hence
\begin{equation}\label{omlip}
\omega = -\alb\left[2\gE+\psi(\gamma)+\psi(1-\gamma)\right]\;,
\end{equation}
$\psi$ being the digamma function and $\gE=-\psi(1)$ the Euler
constant. The solution of eq.~(\ref{omlip})
is $\gamma=\glip$, the Lipatov anomalous dimension:
\begin{equation}\label{glip}
\glip= \alom+
2\zeta(3)\left(\alom\right)^4+
2\zeta(5)\left(\alom\right)^6+
12[\zeta(3)]^2\left(\alom\right)^7+\ldots\;.
\end{equation}
The integrated gluon structure function at scale $Q^2$ is then given by
\begin{equation}\label{FomQ}
 F_\om(Q^2,\mu^2) = 1+\pi\int_{\mu^2}^{Q^2}dk^2\,f_\om(k^2,\mu^2)
=\left(\frac{Q^2}{\mu^2}\right)^{\glip}\;.
\end{equation}

Since we are interested in final states, we shall need to decompose
the structure function into terms corresponding to different numbers
of emitted gluons. The contribution to $f(x,k^2,\mu^2)$ from emission
of $n$ gluons is obtained by iteration of eq.~(\ref{fxk2}):
\begin{equation}
f^{(n)}(x,k^2,\mu^2) = \prod_{i=1}^n\int_{\mu^2}
\frac{dq_i^2}{q_i^2}\frac{d\phi_i}{2\pi}\frac{dz_i}{z_i}
\alb\Delta(z_i,k_i^2,\mu^2)\delta(x-x_n)\delta^2(k-k_n)\;,
\end{equation}
where
\begin{equation}
x_i=\prod_{l=1}^i z_l\;,\>\>\>\>k_i=-\sum_{l=1}^i q_l\;.
\end{equation}
The contribution to the structure function at scale $Q$ is then
obtained by integrating over all $\mu^2<q_i^2<Q^2$:
\begin{equation}\label{Fn}
F^{(n)}(x,Q^2,\mu^2) = \prod_{i=1}^n\int_{\mu^2}^{Q^2}
\frac{dq_i^2}{q_i^2}\frac{d\phi_i}{2\pi}
\frac{dz_i}{z_i}\alb\Delta(z_i,k_i^2,\mu^2)\delta(x-x_n)\;,
\end{equation}
or in terms of the Mellin transform
\begin{equation}\label{Fomn}
F^{(n)}_\om(Q^2,\mu^2) = \prod_{i=1}^n\int_{\mu^2}^{Q^2}
\frac{dq_i^2}{q_i^2}\frac{d\phi_i}{2\pi}H_\om(k_i^2,\mu^2)\;.
\end{equation}

\section{Jet rates}\label{sec_rates}
\subsection{Single-jet rate}\label{sec_1jet}
Consider first the effect of requiring one emitted gluon, say the $j$th,
to have $q_j^2>\mR^2$ while all the others have $q_i^2<\mR^2$.  This defines
the contribution of one resolved gluon plus $n-1$ unresolved,
$F^{(n,1\res)}$:
\begin{equation}
F_\om^{(n,1\res)}(Q^2,\mR^2,\mu^2)= \sum_{j=1}^n
\int_{\mR^2}^{Q^2}\frac{dq_j^2}{q_j^2}\frac{d\phi_j}{2\pi}
H_\om(k_j^2,\mu^2)\prod_{i\neq j}^n\int_{\mu^2}^{\mR^2}
\frac{dq_i^2}{q_i^2}\frac{d\phi_i}{2\pi}H_\om(k_i^2,\mu^2)\;.
\end{equation}
Notice that for $i<j$ the contribution is identical to the $(j-1)$-gluon
contribution to the structure function evaluated at
$Q^2=\mR^2$. On the other hand for $i>j$ we have $k_i^2\simeq q_j^2>\mR^2$.
As shown in the appendix, when $q_j^2>\mR^2$ we can write for any function
$f$, to logarithmic accuracy,
\begin{equation}\label{Hint1}
\int_{\mu^2}^{\mR^2}\frac{dq_i^2}{q_i^2}\frac{d\phi_i}{2\pi}
f(|q_i+q_j|^2) = \ln\left(\frac{\mR^2}{\mu^2}\right)\,f(q_j^2)\;.
\end{equation}
Thus the $q_i$ integrations for $i>j$ become trivial and
\begin{equation}\label{Fn1resom}
F^{(n,1\res)}_\om(Q^2,\mR^2,\mu^2) = \frac{1}{S}
\sum_{j=1}^n F^{(j-1)}_\om(\mR^2,\mu^2)
\int_{\mR^2}^{Q^2}\frac{dq_j^2}{q_j^2}
\left[S\,H_\om(q_j^2,\mu^2)\right]^{n-j+1}
\end{equation}
where we define\footnote{Note that we define
$S$ and $T$ differently from
Refs.~\cite{FSV,Web98} (twice as large) in order to simplify
expressions for the single-logarithmic terms.}
\begin{equation}\label{STdef}
S=\ln(\mR^2/\mu^2)\;,\>\>\>\>T=\ln(Q^2/\mR^2)\;.
\end{equation}
Summing over all $j$ and $n$ gives the total one-jet
contribution,
\begin{eqnarray}\label{F1res}
F^{(1\res)}_\om(Q^2,\mR^2,\mu^2) &=& F_\om(\mR^2,\mu^2)\,
\int_{\mR^2}^{Q^2}\frac{dq_j^2}{q_j^2}
H_\om(q_j^2,\mR^2)\nonumber\\
&=&\exp[\glip S]\,G_\om^{(1)}(T)
\end{eqnarray}
where
\begin{equation}\label{G1def}
G_\om^{(1)}(T)=\ln\left(1+\alom T\right)\;.
\end{equation}
Notice that the collinear-divergent part (the $S$-dependence)
factorizes, and the fraction of events with
one jet is given by the cutoff-independent function
\begin{equation}\label{R1res}
R^{(1\res)}_\om(Q^2,\mR^2) =
\frac{F^{(1\res)}_\om(Q^2,\mR^2,\mu^2)}{F_\om(Q^2,\mu^2)}
\>=\>\exp[-\glip T]\,G_\om^{(1)}(T)\;.
\end{equation}
Thus in the case of the single-jet rate, the only subleading logarithms
are those generated by the presence of the full Lipatov anomalous dimension
(\ref{glip}) in eq.~(\ref{R1res}).

To obtain the jet cross section as a function of $x$, we note
that eq.~(\ref{F1res}) implies that
\begin{equation}\label{F1FG}
F^{(1\res)}_\om(Q^2,\mR^2) = F_\om(\mR^2)\, G_\om^{(1)}(T)\;,
\end{equation}
where we have used the factorization property to replace the
cutoff-dependent gluon structure function $F_\om(\mR^2,\mu^2)$ by
the measured structure function of the target hadron at
scale $\mR^2$, $F_\om(\mR^2)$. It follows that the single-jet
contribution as a function of $x$ is given by the convolution
\begin{equation}\label{F1x}
F^{(1\res)}(x,Q^2,\mR^2) = F(x,\mR^2)\otimes G^{(1)}(x,T)
\equiv\int_x^1\frac{dz}{z} F(z,\mR^2)\otimes G^{(1)}(x/z,T)
\end{equation}
where the inverse Mellin transformation (\ref{melinv}) applied
to eq.~(\ref{G1def}) gives
\begin{equation}\label{G1x}
G^{(1)}(x,T)= \frac{1}{2\pi i}\int_C d\om\,x^{-\om-1}\,G_\om^{(1)}(T)
\>=\>\frac{1-x^{\alb T}}{x\ln(1/x)}\;.
\end{equation}

\subsection{Two-jet rate}\label{sec_2jet}
Now suppose we resolve two gluons $j,j'$ ($j<j'$) with transverse momenta
$q_j^2,q_{j'}^2>\mR^2$. In place of eq.~(\ref{F1res}) we have
\begin{equation}\label{F2res}
F^{(2\res)}_\om(Q^2,\mR^2,\mu^2) = F_\om(\mR^2,\mu^2)\,
\int_{\mR^2}^{Q^2}\frac{dq_j^2}{q_j^2}H_\om(q_j^2,\mR^2)
\int_{\mR^2}^{Q^2}\frac{dq_{j'}^2}{q_{j'}^2}
\frac{d\phi_{j'}}{2\pi}K_\om(|q_j+q_{j'}|^2,\mR^2)
\end{equation}
where, defining the dijet transverse momentum $q_J=q_j+q_{j'}$,
\begin{equation}
K_\om(q_J^2,\mR^2)=
H_\om(q_J^2,\mu^2)\left(1+\sum_{n=j'+1}^\infty
\prod_{i=j'+1}^n\int_{\mu^2}^{\mR^2}\frac{dq_i^2}{q_i^2}
\frac{d\phi_i}{2\pi}H_\om(k_i^2,\mu^2)\right)\;.
\end{equation}
When $q_J^2>\mR^2$ we can safely set $k_i^2=q_J^2$ for $i>j'$,
to obtain
\begin{equation}
K_\om(q_J^2,\mR^2)=H_\om(q_J^2,\mR^2)\;.
\end{equation}
However, this cannot be correct for
$q_J^2<\mR^2$, because $K_\om(q_J^2,\mR^2)$ would then be
infinite at $q_J^2=\mR^2\exp(-\om/\alb)$. Since
$K_\om(q_J^2,\mR^2)$ must be independent of the cutoff
$\mu$, we can evaluate it for $q_J^2<\mR^2$  
by setting $\mu^2=q_J^2$, which gives
\begin{equation}\label{HqltR}
K_\om(q_J^2,\mR^2)=\alom F_\om(\mR^2,q_J^2)
=\alom\left(\frac{\mR^2}{q_J^2}\right)^{\glip}\;.
\end{equation}
Thus we can define the continuous function
\begin{equation}\label{Kom}
K_\om(q_J^2,\mR^2)=H_\om(q_J^2,\mR^2)\theta(q_J^2-\mR^2)
+\alom F_\om(\mR^2,q_J^2)\theta(\mR^2-q_J^2)\;.
\end{equation}
The two-jet rate is then given by
\begin{equation}\label{R2res}
R^{(2\res)}_\om(Q^2,\mR^2) =
\frac{F^{(2\res)}_\om(Q^2,\mR^2,\mu^2)}{F_\om(Q^2,\mu^2)}
\>=\>\exp[-\glip T]\,G_\om^{(2)}(T)
\end{equation}
where
\begin{equation}\label{G2def}
G_\om^{(2)}(T)
=\int_{\mR^2}^{Q^2}\frac{dq_j^2}{q_j^2}\frac{dq_{j'}^2}{q_{j'}^2}
\frac{d\phi_{j'}}{2\pi}K_\om(q_j^2,\mR^2)\,K_\om(|q_j+q_{j'}|^2,\mR^2)\;.
\end{equation}

The integrations in eq.~(\ref{G2def}) can be performed numerically,
without encountering any non-integrable divergences or discontinuities
in the integrand.  The resulting two-jet rate is shown by the solid curve
in fig.~\ref{fig_R2a} for a relatively small value of $\alb/\om$ (0.2),
and for a larger value (0.4) in fig.~\ref{fig_R2b}.
The dashed curves show the result of using the DL prediction
for $G_\om^{(2)}$ in eq.~(\ref{R2res}), i.e.
\begin{equation}\label{R2DL}
R^{(2\res)}_\om(Q^2,\mR^2) \simeq
\exp[-\glip T]\,G_\om^{(2,\DL)}(T)
\end{equation}
where \cite{Web98}
\begin{equation}\label{G2DL}
G_\om^{(2,\DL)}(T) = \frac{1}{2}\ln^2\left(1+\alom T\right)
+\ln\left(1+\alom T\right)
-\alom T\left(1+\alom T\right)^{-1}\;.
\end{equation}
We see that for $\alb/\om=0.2$ the single-logarithmic correction is small,
while for the larger value it is substantial.

\FIGURE{\begin{picture}(0,0)%
\epsfig{file=full2.pstex}%
\end{picture}%
\setlength{\unitlength}{3947sp}%
\begingroup\makeatletter\ifx\SetFigFont\undefined%
\gdef\SetFigFont#1#2#3#4#5{%
  \reset@font\fontsize{#1}{#2pt}%
  \fontfamily{#3}\fontseries{#4}\fontshape{#5}%
  \selectfont}%
\fi\endgroup%
\begin{picture}(5840,3433)(-422,-2999)
\put(795,-2627){\makebox(0,0)[rb]{\smash{\SetFigFont{10}{12.0}{\familydefault}{\mddefault}{\updefault}0}}}
\put(795,-1888){\makebox(0,0)[rb]{\smash{\SetFigFont{10}{12.0}{\familydefault}{\mddefault}{\updefault}0.05}}}
\put(795,-1149){\makebox(0,0)[rb]{\smash{\SetFigFont{10}{12.0}{\familydefault}{\mddefault}{\updefault}0.1}}}
\put(795,-410){\makebox(0,0)[rb]{\smash{\SetFigFont{10}{12.0}{\familydefault}{\mddefault}{\updefault}0.15}}}
\put(795,329){\makebox(0,0)[rb]{\smash{\SetFigFont{10}{12.0}{\familydefault}{\mddefault}{\updefault}0.2}}}
\put(869,-2751){\makebox(0,0)[b]{\smash{\SetFigFont{10}{12.0}{\familydefault}{\mddefault}{\updefault}0}}}
\put(1474,-2751){\makebox(0,0)[b]{\smash{\SetFigFont{10}{12.0}{\familydefault}{\mddefault}{\updefault}2}}}
\put(2079,-2751){\makebox(0,0)[b]{\smash{\SetFigFont{10}{12.0}{\familydefault}{\mddefault}{\updefault}4}}}
\put(2684,-2751){\makebox(0,0)[b]{\smash{\SetFigFont{10}{12.0}{\familydefault}{\mddefault}{\updefault}6}}}
\put(3289,-2751){\makebox(0,0)[b]{\smash{\SetFigFont{10}{12.0}{\familydefault}{\mddefault}{\updefault}8}}}
\put(3894,-2751){\makebox(0,0)[b]{\smash{\SetFigFont{10}{12.0}{\familydefault}{\mddefault}{\updefault}10}}}
\put(4499,-2751){\makebox(0,0)[b]{\smash{\SetFigFont{10}{12.0}{\familydefault}{\mddefault}{\updefault}12}}}
\put(5104,-2751){\makebox(0,0)[b]{\smash{\SetFigFont{10}{12.0}{\familydefault}{\mddefault}{\updefault}14}}}
\put(3137,-2999){\makebox(0,0)[b]{\smash{\SetFigFont{10}{12.0}{\familydefault}{\mddefault}{\updefault}$T=\ln(Q^2/\mu_R^2)$}}}
\put(301,-1149){\makebox(0,0)[b]{\smash{\SetFigFont{10}{12.0}{\familydefault}{\mddefault}{\updefault}$R_\omega^{(2 \res)}$}}}
\end{picture}

\caption{\label{fig_R2a}
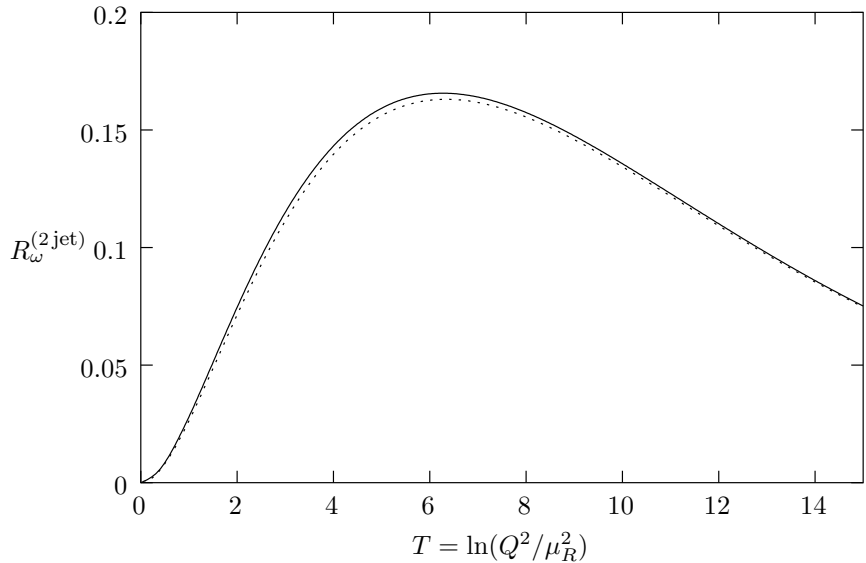
Two-jet rate for $\alb/\om = 0.2$. Dashed: double-log
approximation. Solid: with single-log corrections.}}
\FIGURE{\begin{picture}(0,0)%
\epsfig{file=full4.pstex}%
\end{picture}%
\setlength{\unitlength}{3947sp}%
\begingroup\makeatletter\ifx\SetFigFont\undefined%
\gdef\SetFigFont#1#2#3#4#5{%
  \reset@font\fontsize{#1}{#2pt}%
  \fontfamily{#3}\fontseries{#4}\fontshape{#5}%
  \selectfont}%
\fi\endgroup%
\begin{picture}(5904,3433)(-422,-2999)
\put(795,-2627){\makebox(0,0)[rb]{\smash{\SetFigFont{10}{12.0}{\familydefault}{\mddefault}{\updefault}0}}}
\put(795,-1888){\makebox(0,0)[rb]{\smash{\SetFigFont{10}{12.0}{\familydefault}{\mddefault}{\updefault}0.05}}}
\put(795,-1149){\makebox(0,0)[rb]{\smash{\SetFigFont{10}{12.0}{\familydefault}{\mddefault}{\updefault}0.1}}}
\put(795,-410){\makebox(0,0)[rb]{\smash{\SetFigFont{10}{12.0}{\familydefault}{\mddefault}{\updefault}0.15}}}
\put(795,329){\makebox(0,0)[rb]{\smash{\SetFigFont{10}{12.0}{\familydefault}{\mddefault}{\updefault}0.2}}}
\put(869,-2751){\makebox(0,0)[b]{\smash{\SetFigFont{10}{12.0}{\familydefault}{\mddefault}{\updefault}0}}}
\put(1776,-2751){\makebox(0,0)[b]{\smash{\SetFigFont{10}{12.0}{\familydefault}{\mddefault}{\updefault}2}}}
\put(2684,-2751){\makebox(0,0)[b]{\smash{\SetFigFont{10}{12.0}{\familydefault}{\mddefault}{\updefault}4}}}
\put(3591,-2751){\makebox(0,0)[b]{\smash{\SetFigFont{10}{12.0}{\familydefault}{\mddefault}{\updefault}6}}}
\put(4499,-2751){\makebox(0,0)[b]{\smash{\SetFigFont{10}{12.0}{\familydefault}{\mddefault}{\updefault}8}}}
\put(5406,-2751){\makebox(0,0)[b]{\smash{\SetFigFont{10}{12.0}{\familydefault}{\mddefault}{\updefault}10}}}
\put(301,-1149){\makebox(0,0)[b]{\smash{\SetFigFont{10}{12.0}{\familydefault}{\mddefault}{\updefault}$R_\omega^{(2 \res)}$}}}
\put(3137,-2999){\makebox(0,0)[b]{\smash{\SetFigFont{10}{12.0}{\familydefault}{\mddefault}{\updefault}$T=\ln(Q^2/\mu_R^2)$}}}
\end{picture}

\caption{\label{fig_R2b}
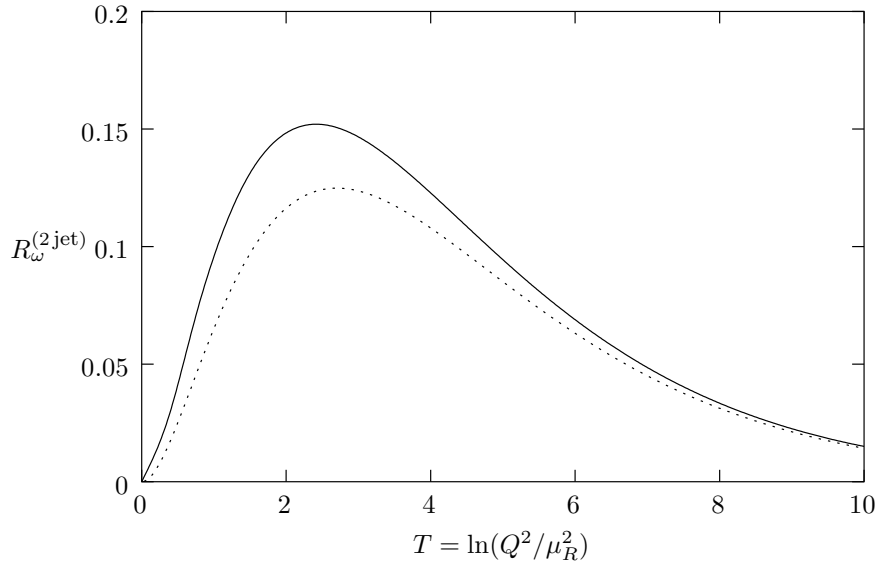
Two-jet rate for $\alb/\om = 0.4$. Dashed: double-log
approximation. Solid: with single-log corrections.}}

Next we consider the perturbative expansion of the two-jet rate.
We can use eq.~(\ref{Kom}) and the results in the appendix to obtain
\begin{eqnarray}\label{Kint}
\int_{\mR^2}^{Q^2}\frac{dq_{j'}^2}{q_{j'}^2}\frac{d\phi_{j'}}{2\pi}
K_\om(|q_j+q_{j'}|^2,\mR^2)&=& 
\int_{\mR^2}^{Q^2}\frac{dq_{j'}^2}{q_{j'}^2}
H_\om(\max\{q_j^2,q_{j'}^2\},\mR^2)\nonumber\\
&+&2\sum_{m=1}^\infty(2m)!\,\zeta(2m+1)
\,[H_\om(q_j^2,\mR^2)]^{2m+1}.
\end{eqnarray}
Substituting in eq.~(\ref{G2def}) we find
\begin{equation}\label{G2res}
G_\om^{(2)}(T) = G_\om^{(2,\DL)}(T) + G_\om^{(2,\SL)}(T)
\end{equation}
where $G_\om^{(2,\DL)}$ is the double-logarithmic result (\ref{G2DL})
and the single-logarithmic correction is
\begin{equation}\label{G2SL}
G_\om^{(2,\SL)}(T) =
2\sum_{m=1}^\infty\frac{(2m)!}{2m+1}\zeta(2m+1)
\left(\alom\right)^{2m+1}
\left[1-\left(1+\alom T\right)^{-2m-1}\right]\;.
\end{equation}

Note that the series in eq.~(\ref{G2SL}) is strongly divergent
for any value of $\alb/\om$.  This is due to the singularity of
$H_\om(q_J^2,\mR^2)$ at $q_J^2=\mR^2\exp(-\om/\alb)<\mR^2$.
The change in $K_\om(q_J^2,\mR^2)$ when $q_J^2<\mR^2$
(see eq.~(\ref{Kom})) removes
the singularity from the integrand in eq.~(\ref{G2def}) but this
does not affect the region of convergence of the perturbation series.
The situation is analogous to the way in which the running of the
QCD coupling $\as(q^2)$ at low values of $q^2$ produces infrared
renormalons \cite{renorm}:
the Landau singularity in the perturbative expression for $\as(q^2)$
leads to a factorial divergence of perturbative expansions with
respect to $\as(Q^2)$, where $Q^2$ is fixed and large, even if
we remove the Landau singularity by making a non-perturbative
modification of $\as(q^2)$ at low $q^2$ \cite{alfamodels}.

In the case of eq.~(\ref{Kint}), the correction arising from the
more careful treatment of the region $q_J^2<\mR^2$, corresponding to
the second term in eq.~(\ref{Kom}), would be
\begin{eqnarray}\label{Hintapp}
\delta\int_{\mR^2}^{Q^2}\frac{dq_{j'}^2}{q_{j'}^2}
\frac{d\phi_{j'}}{2\pi} K_\om(q_J^2,\mR^2)
&\simeq&
\alom\int_0^{\mR^2}\frac{dq_J^2}{q_j^2}
\left[\left(\frac{\mR^2}{q_J^2}\right)^{\glip}
-\sum_{m=0}^\infty\left(-\alom
\ln\frac{q_J^2}{\mR^2}\right)^m\right]\nonumber\\
&=&\alom\frac{\mR^2}{q_j^2}
\left[\frac{1}{1-\glip}
-\sum_{m=0}^\infty m!\left(\alom\right)^m\right]\;.
\end{eqnarray}
This does indeed contain a factorially divergent series, but,
owing to the overall factor of $1/q_j^2$, it does not contribute
any logarithms of $Q^2/\mR^2$ upon substitution in eq.~(\ref{G2def}).
Therefore a more accurate treatment of the region $q_J^2<\mR^2$,
although necessary to evaluate the integrals,
does not affect the two-jet rate to SL precision.

If we interpret the series in eq.~(\ref{G2SL})
as an asymptotic expansion, then the partial sum truncated
after the smallest term represents an estimate of the total SL
correction, with an uncertainty of the order of the smallest term.
This estimate is shown by the points in figs.~\ref{fig_dR2a} and
\ref{fig_dR2b}, with the uncertainty represented by the
error bars.  We see that estimates from the series are of the
same order of magnitude as the numerical results, but the
discrepancy may be several times the expected uncertainty.
At the smaller value of $\alb/\om$, the SL correction is
relatively small (c.f.\ fig.~\ref{fig_R2a}) and the discrepancy is
not so important. For the larger value of $\alb/\om$, the estimate
is better than expected, but the correction and the uncertainty
are both large.

\FIGURE{\begin{picture}(0,0)%
\epsfig{file=corr2.pstex}%
\end{picture}%
\setlength{\unitlength}{3947sp}%
\begingroup\makeatletter\ifx\SetFigFont\undefined%
\gdef\SetFigFont#1#2#3#4#5{%
  \reset@font\fontsize{#1}{#2pt}%
  \fontfamily{#3}\fontseries{#4}\fontshape{#5}%
  \selectfont}%
\fi\endgroup%
\begin{picture}(6284,3433)(-770,-2999)
\put(869,-2627){\makebox(0,0)[rb]{\smash{\SetFigFont{10}{12.0}{\familydefault}{\mddefault}{\updefault}0}}}
\put(869,-2134){\makebox(0,0)[rb]{\smash{\SetFigFont{10}{12.0}{\familydefault}{\mddefault}{\updefault}0.002}}}
\put(869,-1642){\makebox(0,0)[rb]{\smash{\SetFigFont{10}{12.0}{\familydefault}{\mddefault}{\updefault}0.004}}}
\put(869,-1149){\makebox(0,0)[rb]{\smash{\SetFigFont{10}{12.0}{\familydefault}{\mddefault}{\updefault}0.006}}}
\put(869,-656){\makebox(0,0)[rb]{\smash{\SetFigFont{10}{12.0}{\familydefault}{\mddefault}{\updefault}0.008}}}
\put(869,-164){\makebox(0,0)[rb]{\smash{\SetFigFont{10}{12.0}{\familydefault}{\mddefault}{\updefault}0.01}}}
\put(869,329){\makebox(0,0)[rb]{\smash{\SetFigFont{10}{12.0}{\familydefault}{\mddefault}{\updefault}0.012}}}
\put(943,-2751){\makebox(0,0)[b]{\smash{\SetFigFont{10}{12.0}{\familydefault}{\mddefault}{\updefault}0}}}
\put(1548,-2751){\makebox(0,0)[b]{\smash{\SetFigFont{10}{12.0}{\familydefault}{\mddefault}{\updefault}2}}}
\put(2153,-2751){\makebox(0,0)[b]{\smash{\SetFigFont{10}{12.0}{\familydefault}{\mddefault}{\updefault}4}}}
\put(2758,-2751){\makebox(0,0)[b]{\smash{\SetFigFont{10}{12.0}{\familydefault}{\mddefault}{\updefault}6}}}
\put(3363,-2751){\makebox(0,0)[b]{\smash{\SetFigFont{10}{12.0}{\familydefault}{\mddefault}{\updefault}8}}}
\put(3968,-2751){\makebox(0,0)[b]{\smash{\SetFigFont{10}{12.0}{\familydefault}{\mddefault}{\updefault}10}}}
\put(4573,-2751){\makebox(0,0)[b]{\smash{\SetFigFont{10}{12.0}{\familydefault}{\mddefault}{\updefault}12}}}
\put(5178,-2751){\makebox(0,0)[b]{\smash{\SetFigFont{10}{12.0}{\familydefault}{\mddefault}{\updefault}14}}}
\put(153,-1149){\makebox(0,0)[b]{\smash{\SetFigFont{10}{12.0}{\familydefault}{\mddefault}{\updefault}$\delta R_\omega^{(2 \res)}$}}}
\put(3211,-2999){\makebox(0,0)[b]{\smash{\SetFigFont{10}{12.0}{\familydefault}{\mddefault}{\updefault}$T=\ln(Q^2/\mu_R^2)$}}}
\end{picture}

\caption{\label{fig_dR2a}
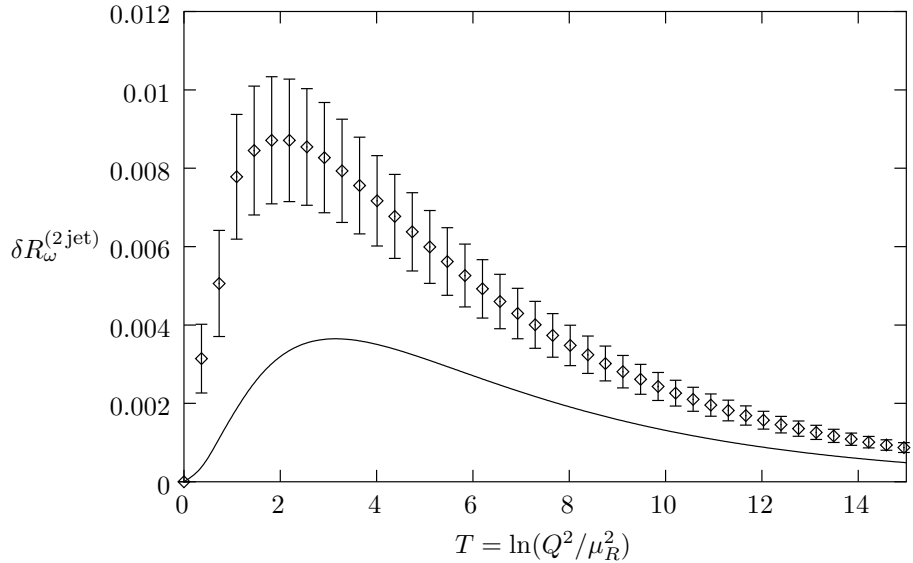
Solid: single-log correction to two-jet rate for $\alb/\om = 0.2$.
Points: estimate from asymptotic expansion; `error bars'
indicate the smallest term in the expansion.}}
\FIGURE{\begin{picture}(0,0)%
\epsfig{file=corr4.pstex}%
\end{picture}%
\setlength{\unitlength}{3947sp}%
\begingroup\makeatletter\ifx\SetFigFont\undefined%
\gdef\SetFigFont#1#2#3#4#5{%
  \reset@font\fontsize{#1}{#2pt}%
  \fontfamily{#3}\fontseries{#4}\fontshape{#5}%
  \selectfont}%
\fi\endgroup%
\begin{picture}(6252,3433)(-770,-2999)
\put(795,-2627){\makebox(0,0)[rb]{\smash{\SetFigFont{10}{12.0}{\familydefault}{\mddefault}{\updefault}0}}}
\put(795,-2036){\makebox(0,0)[rb]{\smash{\SetFigFont{10}{12.0}{\familydefault}{\mddefault}{\updefault}0.02}}}
\put(795,-1445){\makebox(0,0)[rb]{\smash{\SetFigFont{10}{12.0}{\familydefault}{\mddefault}{\updefault}0.04}}}
\put(795,-853){\makebox(0,0)[rb]{\smash{\SetFigFont{10}{12.0}{\familydefault}{\mddefault}{\updefault}0.06}}}
\put(795,-262){\makebox(0,0)[rb]{\smash{\SetFigFont{10}{12.0}{\familydefault}{\mddefault}{\updefault}0.08}}}
\put(795,329){\makebox(0,0)[rb]{\smash{\SetFigFont{10}{12.0}{\familydefault}{\mddefault}{\updefault}0.1}}}
\put(869,-2751){\makebox(0,0)[b]{\smash{\SetFigFont{10}{12.0}{\familydefault}{\mddefault}{\updefault}0}}}
\put(1776,-2751){\makebox(0,0)[b]{\smash{\SetFigFont{10}{12.0}{\familydefault}{\mddefault}{\updefault}2}}}
\put(2684,-2751){\makebox(0,0)[b]{\smash{\SetFigFont{10}{12.0}{\familydefault}{\mddefault}{\updefault}4}}}
\put(3591,-2751){\makebox(0,0)[b]{\smash{\SetFigFont{10}{12.0}{\familydefault}{\mddefault}{\updefault}6}}}
\put(4499,-2751){\makebox(0,0)[b]{\smash{\SetFigFont{10}{12.0}{\familydefault}{\mddefault}{\updefault}8}}}
\put(5406,-2751){\makebox(0,0)[b]{\smash{\SetFigFont{10}{12.0}{\familydefault}{\mddefault}{\updefault}10}}}
\put(153,-1149){\makebox(0,0)[b]{\smash{\SetFigFont{10}{12.0}{\familydefault}{\mddefault}{\updefault}$\delta R_\omega^{(2 \res)}$}}}
\put(3137,-2999){\makebox(0,0)[b]{\smash{\SetFigFont{10}{12.0}{\familydefault}{\mddefault}{\updefault}$T=\ln(Q^2/\mu_R^2)$}}}
\end{picture}

\caption{\label{fig_dR2b}
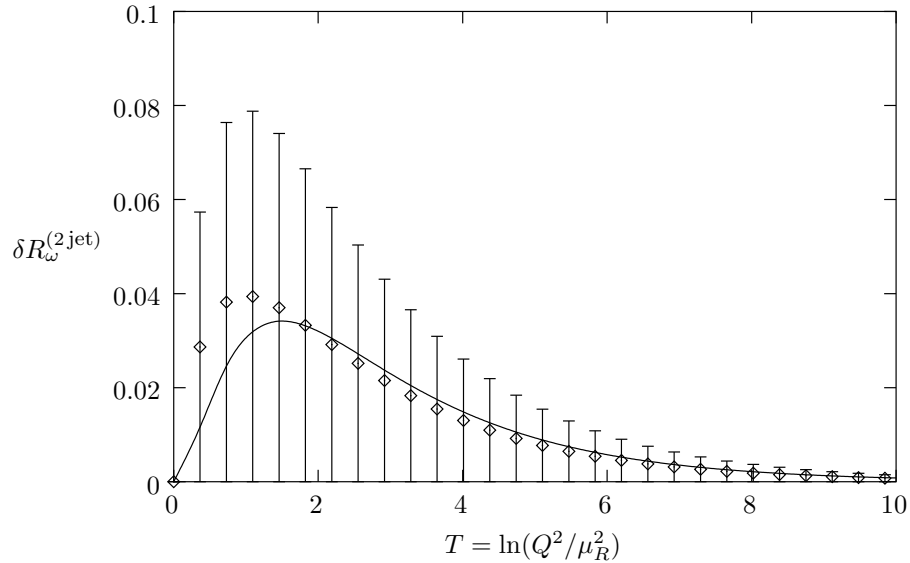
Solid: single-log correction to two-jet rate for $\alb/\om = 0.4$.
Points: estimate from asymptotic expansion; `error bars'
indicate the smallest term in the expansion.}}

To deduce the two-jet rate as a function of $x$, we can proceed
as in eq.~(\ref{F1x}), writing
\begin{equation}\label{F2x}
F^{(2\res)}(x,Q^2,\mR^2) = F(x,\mR^2)\otimes G^{(2)}(x,T)
\end{equation}
where
\begin{equation}\label{G2x}
G^{(2)}(x,T) = \frac{1}{2\pi i}\int_C d\om\,x^{-\om-1}
\,G_\om^{(2)}(T)\;.
\end{equation}
For the DL contribution we find from eq.~(\ref{G2DL}) that
\begin{equation}\label{G2DLx}
G^{(2,\DL)}(x,T) = \frac{1}{x\ln(1/x)}\,\cG^{(2,\DL)}[\alb T\ln(1/x)]
\end{equation}
where
\begin{equation}\label{cG2z}
\cG^{(2,\DL)}[z] = E_1(z)+\ln z +\gE+1
+e^{-z}\left[E_1(-z)+\ln z +\gE-1-z\right]\;,
\end{equation}
$E_1(z)$ being the exponential integral function
\begin{equation}\label{E1z}
E_1(z) = \int_z^\infty dt\frac{e^{-t}}{t}\;,
\end{equation}
interpreted as a principal-value integral when $z<0$.

The divergence of the series for the SL correction in $\om$-space,
eq.~(\ref{G2SL}), is cured when one makes the inverse Mellin
transformation to $x$-space, because the factorial coefficients
are cancelled:
\begin{equation}\label{melom}
\frac{1}{2\pi i}\int_C d\om\,x^{-\om-1}\left(\alom\right)^{2m+1}
=\frac{\alb}{x}\frac{[\alb\ln(1/x)]^{2m}}{(2m)!}\;.
\end{equation}
Thus the SL correction to the two-jet rate can be expressed in closed
form as a function of $x$:
\begin{equation}\label{G2SLx}
G^{(2,\SL)}(x,T) = \frac{1-x^{\alb T}}{x\ln(1/x)}
\,\cG^{(2,\SL)}[\alb\ln(1/x)]
\end{equation}
where
\begin{equation}\label{cG2ans}
\cG^{(2,\SL)}[z] = \ln\Gamma(1-z)-\ln\Gamma(1+z)-2\gE z\;.
\end{equation}
Notice that the expression (\ref{G2SLx}) is singular at $\alb\ln(1/x)=1$.
From the viewpoint of the Mellin transformation (\ref{mellin}),
it is this singularity that produces the divergence of the series in
$\om$-space. Conversely, use of the more correct expression (\ref{G2def})
in $\om$-space, with the kernel function $K_\om$ given in eq.~(\ref{Kom}),
should suffice to remove the singularity in $x$-space.

\subsection{Three-jet rate}\label{sec_3jet}
The method used above for the two-jet rate can be extended, albeit
laboriously, to higher jet multiplicities. In the three-jet case we have
\begin{equation}\label{R3res}
R^{(3\res)}_\om(Q^2,\mR^2) =
\frac{F^{(3\res)}_\om(Q^2,\mR^2,\mu^2)}{F_\om(Q^2,\mu^2)}
\>=\>\exp[-\glip T]\,G_\om^{(3)}(T)
\end{equation}
where
\begin{equation}\label{G3def}
G_\om^{(3)}(T) =\int_{\mR^2}^{Q^2}\frac{dq_j^2}{q_j^2}
\frac{dq_{j'}^2}{q_{j'}^2}\frac{dq_{j''}^2}{q_{j''}^2}
\frac{d\phi_{j'}}{2\pi}\frac{d\phi_{j''}}{2\pi}
K_\om(q_j^2,\mR^2) K_\om(|q_j+q_{j'}|^2,\mR^2)
K_\om(|q_j+q_{j'}+q_{j''}|^2,\mR^2).
\end{equation}
One could in principle evaluate
this expression numerically using eq.~(\ref{Kom}) for $K_\om$.
Here we derive the perturbative expansion analogous to eq.~(\ref{G2SL}).
Introducing $t=\ln(q_j^2/\mR^2)$ etc.\ for brevity, the results in
the appendix give
\begin{equation}
G_\om^{(3)}(T) =
\int_0^T dt\,H_\om(t)\left[\int_0^T dt'\,L_\om\left(\max\{t,t'\}\right)
+2\sum_{m=1}^\infty\zeta(2m+1)\,
\frac{\partial^{2m}L_\om}{\partial t^{2m}}\right]
\end{equation}
where we write $H_\om(q_j^2,\mR^2)$ as $H_\om(t)$ and
\begin{equation}
L_\om(t)= H_\om(t)[\ln H_\om(t) - \ln H_\om(T)] + t\,[H_\om(t)]^2
+2\sum_{m'=1}^\infty(2m')!\,\zeta(2m'+1)\,[H_\om(t)]^{2m'+2}.
\end{equation}
Hence we obtain
\begin{equation}\label{G3res}
G_\om^{(3)}(T)=G_\om^{(3,\DL)}(T)+G_\om^{(3,\SL)}(T)
\end{equation}
where $G_\om^{(3,\DL)}$ is the double-logarithmic result \cite{Web98}
\begin{eqnarray}
&&G_\om^{(3,\DL)}(T)\>=\>
\frac{1}{6}\ln^3\left(1+\alom T\right)
+\ln^2\left(1+\alom T\right)\nonumber\\
&&+\ln\left(1+\alom T\right)
\left(1+\alom T\right)^{-1}
-\alom T\left(1+\frac{3\alb}{2\om}T\right)
\left(1+\alom T\right)^{-2}
\end{eqnarray}
and
\begin{eqnarray}\label{G3SL}
&&G_\om^{(3,\SL)}(T)\>=\>
-2\sum_{m=1}^\infty (2m)!\,\zeta(2m+1)
\left(\alom\right)^{2m+1}\Biggl\{
\left[1-\left(1+\alom T\right)^{-2m-2}\right]\nonumber\\
&&-\frac{1}{2m+1}\left[\ln\left(1+\alom T\right)
+\psi(2m+1)+\gE+2\right]\left[1-\left(1+
\alom T\right)^{-2m-1}\right]\Biggr\}+\nonumber\\
&&+\,4\sum_{m,m'=1}^\infty \frac{(2m+2m'+1)!}{(2m+2m'+2)(2m'+1)}\,
\zeta(2m+1)\,\zeta(2m'+1)
\left(\alom\right)^{2m+2m'+2}\nonumber\\
&&\times
\left[1-\left(1+\alom T\right)^{-2m-2m'-2}\right]\;.
\end{eqnarray}

The expansion in eq.~(\ref{G3SL}) is again strongly divergent for
all values of $\alb/\om$. As discussed in subsect.~\ref{sec_2jet},
it can still be interpreted as an asymptotic expansion and used
as a guide to the order of magnitude of the SL correction.
Furthermore, the divergence is cured upon inverting the
Mellin transformation to obtain the jet rate in $x$-space,
as long as $\alb\ln(1/x)<1$.  To extend
the prediction to smaller values of $x$ one would need to
evaluate eq.~(\ref{G3def}) numerically using the full
kernel function in eq.~(\ref{Kom}).

\section{Generating function for multi-jet rates}\label{sec_njet}
For a general jet multiplicity $r$, we can write
\begin{equation}\label{Rrres}
R^{(r\res)}_\om(Q^2,\mR^2) = \frac{1}{r!}
\left.\frac{\partial^r}{\partial u^r}R_\om(u,T)\right|_{u=0}\;,
\end{equation}
where the jet-rate generating function $R_\om$ is given by
\begin{equation}
R_\om(u,T)=\exp[-\glip T]\,G_\om(u,T)
\end{equation}
and
\begin{equation}\label{GuT}
G_\om(u,T)=\sum_{r=0}^\infty u^r G^{(r)}_\om(T)\;.
\end{equation}
The function  $G^{(r)}_\om(T)$ was given in eqs.~(\ref{G1def}),
(\ref{G2res}) and (\ref{G3res}) for $r=1,2$ and 3, respectively.

We can obtain the perturbative expansion of the function
$G_\om(u,T)$ as follows.  We first define the unintegrated function
$g_\om(u,t,T)$ such that
\begin{equation}\label{Gexpn}
G_\om(u,T)= 1+\int_0^T dt\,g_\om(u,t,T)\;.
\end{equation}
Then using the results in the appendix we find that $g_\om(u,t,T)$
satisfies the integro-differential equation
\begin{equation}\label{gomu}
g_\om(u,t,T) = uH_\om(t)\left[1+
\int_0^T dt'\,g_\om\left(u,\max\{t,t'\},T\right)
+2\sum_{m=1}^\infty\zeta(2m+1)\,
\frac{\partial^{2m}g_\om}{\partial t^{2m}}\right]\;.
\end{equation}
Writing
\begin{equation}
g_\om(u,t,T) = \sum_{n=0}^\infty c_n(u,t,T)
\left(\alom\right)^n\;,
\end{equation}
this implies that
\begin{eqnarray}
c_{n+1}(u,t,T)&=&u\,\delta_{n,0}-(1-u)t\,c_n(u,t,T)
+u\int_t^T dt'\,c_n(u,t',T)\nonumber\\
&&+2u\sum_{m=1}^\infty\zeta(2m+1)
\frac{\partial^{2m}c_n}{\partial t^{2m}}\,.
\end{eqnarray}
Starting from $c_0=0$, this gives $c_n$ iteratively
as a polynomial in $u$, $t$ and $T$, which can be substituted in
eq.~(\ref{Gexpn}) to obtain the perturbative expansion of $G_\om(u,T)$
to any desired order.\footnote{The results agree with those given
to fourth order in ref.~\cite{M}.} The relation (\ref{melom}) can
then be used to transform the result directly to $x$-space, giving
\begin{equation}\label{Frresx}
F^{(r\res)}(x,Q^2,\mR^2) = \frac{1}{r!}F(x,\mR^2)\otimes
\left.\frac{\partial^r}{\partial u^r}G(u,x,T)\right|_{u=0}\;,
\end{equation}
where
\begin{equation}\label{GuxT}
G(u,x,T)= \delta(1-x)+\int_0^T dt\,g(u,x,t,T)
\end{equation}
with
\begin{equation}
g(u,x,t,T) = \frac{\alb}{x}\sum_{n=0}^\infty c_{n+1}(u,t,T)
\frac{[\alb\ln(1/x)]^n}{n!}\;,
\end{equation}
which we believe to be a convergent series as long as $\alb\ln(1/x)<1$.

\subsection{Anomalous dimension}
Notice that for $u=1$ we have
\begin{equation}
R_\om(1,T)=\sum_{r=0}^\infty R^{(r\res)}_\om(Q^2,\mR^2) = 1
\end{equation}
and therefore
\begin{equation}\label{G1T}
G_\om(1,T)=\exp[\glip T]\;.
\end{equation}
To show that eq.~(\ref{gomu}) does indeed lead to the Lipatov
result (\ref{omlip}) for the anomalous dimension,
we note that when $u=1$ the solution of eq.~(\ref{gomu}) is
\begin{eqnarray}\label{gom1}
g_\om(1,t,T) &=& \gamma\,e^{\gamma(T-t)}\nonumber\\
G_\om(1,T)&=& 1+\int_0^T dt\,g_\om(1,t,T)=e^{\gamma T}
\end{eqnarray}
where
\begin{eqnarray}
\gamma &=& H_\om(t)\left[1+\gamma t
+2\sum_{m=1}^\infty\zeta(2m+1)\,\gamma^{2m+1}\right]\nonumber\\
&=& \alom\left[1+
2\sum_{m=1}^\infty\zeta(2m+1)\,\gamma^{2m+1}\right]\nonumber\\
&=& \alom\gamma\left[\frac{1}{\gamma}-2\gE
-\psi(1+\gamma)-\psi(1-\gamma)\right]\;.
\end{eqnarray}
Rearranging terms, we obtain $\gamma=\glip$
given by eqs.~(\ref{omlip}) and (\ref{glip}).

\subsection{Jet multiplicity moments}\label{sec_mom}
We can compute the moments of the jet multiplicity distribution by
successively differentiating the generating function at $u=1$:
\begin{equation}
\langle r(r-1)\ldots(r-s+1)\rangle =
\left.\frac{\partial^s}{\partial u^s}R_\om(u,T)\right|_{u=1}\;.
\end{equation}
In this way we obtain the perturbative expansion of the mean number
of jets
\begin{eqnarray}
\langle r\rangle &=&
T\alom +\frac{1}{2}T^2\left(\alom\right)^2
+2\zeta(3)T\left(\alom\right)^4\nonumber\\
&&+4\zeta(3)T^2\left(\alom\right)^5
-8\zeta(5)T\left(\alom\right)^6 +\cdots
\end{eqnarray}
and the mean square fluctuation in this number,
\begin{eqnarray}
\langle r^2\rangle -\langle r\rangle^2 &=&
T\alom+\frac{3}{2}T^2\left(\alom\right)^2
+\frac{2}{3}T^3\left(\alom\right)^3
-2\zeta(3)T\left(\alom\right)^4\nonumber\\
&&+12\zeta(3)T^2\left(\alom\right)^5
-\left(8\zeta(5)T -\frac{40}{3}\zeta(3)T^3\right)
\left(\alom\right)^6 +\cdots\;.
\end{eqnarray}

It appears true to all orders to SL precision that, as in the DL
approximation \cite{Web98}, the mean number of jets is a quadratic
function of $T$ and the mean square fluctuation is
a cubic function of $T$. Thus the distribution of jet multiplicity at
small $x$ and large $T$ is narrow, in the sense that its r.m.s.\ width
increases less rapidly than its mean as $T$ increases.

\section{Conclusions}\label{sec_conc}
The calculation of jet rates at small $x$ poses many interesting
challenges and sheds new light on the novel dynamics of this
kinematic region. In the present paper we have concentrated on
those perturbative contributions which have a factor of
$\ln(1/x)$ for each power of $\as$ and are further enhanced
by one or more powers of $T=\ln(Q^2/\mR^2)$, $\mR$ being
the minimum resolved jet transverse momentum.

For sufficiently large values of $\mR^2$ and $Q^2\gg\mR^2$,
the resummation of such terms would seems to be a
well-defined problem in perturbation theory.
The results in sect.~\ref{sec_njet} do indeed specify
all terms of the form $(\alb/\om)^n T^m$ with $m>0$,
for any jet multiplicity, $\om$ being the moment variable
in the Mellin transform.  However, as we have seen explicity
for the two- and three-jet rates (and we believe to be true
more generally), the single-logarithmic terms (those
with $0<m<n$) cannot be resummed directly since
they form strongly divergent series. In $\om$-space, the
divergence is associated with kinematic regions in which the
vector sums of transverse momenta of combinations of jets
are less than $\mR$. A more careful treatment of such regions
renders the jet rates well-defined as integrals. Furthermore,
one obtains convergent series, within a limited range of $x$,
after performing the inverse Mellin transformation to $x$-space.
In the case of the two-jet rate, we were able to sum the
resulting series explicitly, to obtain a closed-form expression
valid in the region $\alb\ln(1/x)<1$.

A number of interesting questions arise from our results.
Clearly one would like to extend the resummation of jet rates
to higher multiplicities and smaller values of $x$. This will
require an $x$-space treatment of the difficult kinematic
regions mentioned above. One would also like to prove the
conjectures in subsect.~\ref{sec_mom} about jet multiplicity
moments to all orders, and preferably to resum them.
Ultimately, next-to-leading terms in $\ln(1/x)$ should
also be included. Such terms will arise from  next-to-leading
corrections to the BFKL kernel and from the resolution of emitted
gluons into two jets.

\acknowledgments
We thank G.\ Salam for valuable comments and especially for
pointing out eq.~(\ref{HqltR}).
BRW is also grateful to  S.\ Catani and G.\ Marchesini
for many helpful discussions.

\appendix
\section{Useful integrals}
To evaluate integrals of the form
\begin{equation}\label{logint2}
\int_{\mu^2}^{Q^2}\frac{dq^2}{q^2}\frac{d\phi}{2\pi}f(|q+k|^2)
\end{equation}
to logarithmic accuracy, we may assume that $f(k^2)$ has an
expansion in powers of $\ln k^2$, and so it suffices to
consider $f(k^2)=\ln^p k^2$ for positive integer values of $p$.
This can in turn be done by considering $f(k^2)= k^{2\nu}$ for
small $\nu$ and extracting the coefficient of $\nu^p/p!$.  Now
\begin{eqnarray}\label{powint}
\int_{\mu^2}^{Q^2}\frac{dq^2}{q^2}\frac{d\phi}{2\pi}|q+k|^{2\nu}
&=&k^{2\nu} \int_{\mu^2}^{Q^2}\frac{dq^2}{q^2}\frac{d\phi}{2\pi}
\left(1+\frac q k e^{i\phi}\right)^\nu
\left(1+\frac q k e^{-i\phi}\right)^\nu \nonumber\\
&=&k^{2\nu} \sum_{m=0}^\infty \combi{\nu}{m}^2
\left[\int_{\mu^2}^{k^2}\frac{dq^2}{q^2}
\left(\frac{q^2}{k^2}\right)^m +
\int_{k^2}^{Q^2}\frac{dq^2}{q^2}
\left(\frac{q^2}{k^2}\right)^{\nu-m}\right]\;.
\end{eqnarray}
If $k^2<\mu^2$ the first integral on the right-hand side
is absent and the lower limit on the second becomes $\mu^2$;
if $k^2>Q^2$ the second integral is absent and the upper
limit on the first becomes $Q^2$.  Thus, neglecting
power-suppressed terms, we have
\begin{eqnarray}\label{powint2}
\int_{\mu^2}^{Q^2}\frac{dq^2}{q^2}\frac{d\phi}{2\pi}|q+k|^{2\nu}
&=&\frac{1}{\nu}\left(Q^{2\nu}-\mu^{2\nu}\right)
\>\>\>\mbox{for $k^2<\mu^2$,}\nonumber\\
&=&k^{2\nu}\Biggl[\ln\left(\frac{k^2}{\mu^2}\right)
+\frac{1}{\nu}\left(\frac{Q^2}{k^2}\right)^\nu-
\frac 1{\nu}\nonumber\\
&&-2\gE-\psi(1+\nu)-\psi(1-\nu)\Biggr]
\>\>\>\mbox{for $\mu^2<k^2<Q^2$,}\nonumber\\
&=&k^{2\nu}\ln\left(\frac{Q^2}{\mu^2}\right)
\>\>\>\mbox{for $k^2>Q^2$.}
\end{eqnarray}
Here we have used the remarkable result, valid for
$\mbox{Re}\;\nu>-1$,
\begin{equation}
\sum_{m=1}^\infty \combi{\nu}{m}^2
\left[\frac 1 m +\frac 1{m-\nu}\right]
=2\sum_{m=1}^\infty\zeta(2m+1)\,\nu^{2m}
=-2\gE-\psi(1+\nu)-\psi(1-\nu)\,.
\end{equation}
Thus, neglecting power-suppressed terms, we find
\begin{eqnarray}
\int_{\mu^2}^{Q^2}\frac{dq^2}{q^2}\frac{d\phi}{2\pi}f(|q+k|^2)
&=&\int_{\mu^2}^{Q^2}\frac{dq^2}{q^2}f(q^2)
\>\>\>\mbox{for $k^2<\mu^2$,}\nonumber\\
&=&\int_{\mu^2}^{Q^2}\frac{dq^2}{q^2}f(\max\{k^2,q^2\})\nonumber\\
&&+2\sum_{m=1}^\infty \zeta(2m+1)
\left(k^2\frac{\partial}{\partial k^2}\right)^{2m}f(k^2)
\>\>\>\mbox{for $\mu^2<k^2<Q^2$,}\nonumber\\
&=& \ln\left(\frac{Q^2}{\mu^2}\right)f(k^2)
\>\>\>\mbox{for $k^2>Q^2$.}
\end{eqnarray}

\end{document}